# Active control of magnetoresistance of organic spin valves using ferroelectricity

Dali Sun[2,3, §, ||], Mei Fang[1, §], Xiaoshan Xu[2, *, †], Lu Jiang[2,3], Hangwen Guo[2,3], Yanmei Wang[1], Wenting Yang[1], Lifeng Yin[1], Paul C. Snijders[2], T. Z. Ward[2], Zheng Gai[2,5], X.-G. Zhang[4,5] Ho Nyung Lee[2], and Jian Shen[1,3, *, ‡]

[1]State Key Laboratory of Surface Physics and Department of Physics, Fudan University, Shanghai 200433, China

[2]Materials Science and Technology Division, Oak Ridge National Laboratory, Oak Ridge, TN 37831, USA

[3]Department of Physics and Astronomy, The University of Tennessee, Knoxville, TN 37996, USA

[4]Computer Science and Mathematics Division, Oak Ridge National Laboratory, Oak Ridge, TN 37831, USA

[5]Center for Nanophase Materials Sciences, Oak Ridge National Laboratory, Oak Ridge, TN 37831, USA

* To whom correspondence should be addressed.

† xiaoshan.xu@unl.edu; ‡ shenj5494@fudan.edu.cn

§D. S and M. F equally contributed to this work.

‖Current address: Department of Physics and Astronomy, University of Utah, Salt Lake City, UT 84112, USA

**Organic spintronic devices have been appealing because of the long spin life time of the charge carriers in the organic materials and their low cost, flexibility and chemical diversity. In previous studies, the control of resistance of organic spin valves is generally achieved by the alignment of the magnetization directions of the two ferromagnetic electrodes, generating magnetoresistance.[1] Here we employ a new knob to tune the resistance of organic spin valves by adding a thin ferroelectric interfacial layer between the ferromagnetic electrode and the organic spacer. We show that the resistance can be controlled by not only the spin alignment of the two ferromagnetic electrodes, but also by the electric polarization of the interfacial ferroelectric layer: the MR of the spin valve depends strongly on the history of the bias voltage which is correlated with the polarization of the ferroelectric layer; the MR even changes sign when the electric polarization of the ferroelectric layer is reversed. This new tunability can be understood in terms of the change of relative energy level alignment between ferromagnetic electrode and the organic spacer caused by the electric dipole moment of the ferroelectric layer. These findings enable active control of resistance using both electric and magnetic fields, opening up possibility for multi-state organic spin valves and shed light on the mechanism of the spin transport in organic spin valves.**

Since the pioneering work by Dediu and Xiong,[2–5] many of the follow-up studies have focused on achieving high magnetoresistance (*MR*) in organic spin valves (OSV) and uncovering the underlying transport mechanisms.[3,4,6–11] It has been generally acknowledged that the interfaces between the organic layer and the ferromagnetic electrodes play a critical role in determining the actual spin-dependent transport mechanism.[12–15] Barraud *et al* studied the spin transport of an OSV in which the thin organic layer serves as a tunneling barrier as opposed to a diffusive spacer.[16] A spin transport model describing the role of interfacial spin-dependent metal/organic hybridization on the amplitude and sign of the MR was put forward.[16,17] Recently, Schulz *et al.* observed a reversal of the spin polarization of extracted charge carriers by introducing a thin interfacial permanent dipolar layer (LiF).[18] This work indicates that the local electric dipole moment at the interface is important for MR, although in such a device the added dipolar layer plays only a passive role in terms of controlling MR.

In order to achieve active control of MR, we employ a ferroelectric (FE) thin layer at the interface in organic spin valves. The material of choice is $PbZr_{0.2}Ti_{0.8}O_3$ (PZT), which has a large polarization (~80 $\mu C/cm^2$)[19]. The PZT layer can induce strong interfacial dipoles and built-in electric field between the organic spacer layer and the ferromagnetic electrode. The interfacial dipole is switchable by external electric field, potentially allowing the control of the spin polarization of injected carriers in organic spintronics.

Our fabricated organic spin valves consist a 65 nm $Alq_3$ (tris-[8-hydroxyquinoline] aluminum)

layer sandwiched between a 5-nm-thick PZT layer epitaxially grown on a 30-nm-thick $La_{0.67}Sr_{0.33}MnO_3$ (LSMO) film buffered $SrTiO_3$ (STO) substrate, and a 10-nm-thick (nominal thickness) cobalt layer with gold capping. In this LSMO/PZT/Alq$_3$/Co junction (FE-OSV), Co and LSMO serve as top and bottom magnetic electrodes, respectively. The device structure is schematically shown in **Fig. 1a**. As discussed in the following, such kind of devices exhibit striking tunability, i.e. both the amplitude and sign of MR are tunable due to the presence of the ferroelectric PZT.

**Surface morphology and the polarization reversal of the PZT films**

After epitaxial growth of PZT on LSMO/STO, the PZT layer has a smooth surface with atomically flat terraces, as characterized by atomic force microscopy (AFM) shown in **Fig. 1b**. This provides an ideal base for preparing a well-defined Alq$_3$/PZT interface. The purpose of introducing the 5-nm-thick-PZT is to tune the energy level alignment between the Alq$_3$ and LSMO layers because the polarization in PZT is switchable by applying an electric field. Therefore, the polarization reversal of the PZT is crucial. Here we have characterized the polarization reversal of the PZT films used in the FE-OSV devices with piezoelectric force microscopy (PFM). **Fig. 1c** shows a PFM image of the PZT film with a part of the film poled by a conducting tip with +/-2.5V relative to the LSMO bottom electrode. It shows that the polarization of the as-grown films is pointing "up", and a clear reversal of polarization between "up" and "down" states can be created by sign reversal of the applied voltage. **Fig. 1d** shows the voltage dependence of the polarization. It is clear that the polarization of the PZT film (5 nm in thickness) starts to switch to the "down" state when the voltage of the scanning probe exceeds 0.8 V. The polarization reversal of the PZT in the FE-OSV is characterized by the measurements of the hysteretic polarization-voltage dependence. As shown in **Fig. 1e**, with the Alq$_3$/Co/Au layers on top of PZT, the coercive voltage is ~ 2 V. **Fig. 1f** shows the Transmission Electron Microscope (TEM) image for the cross section of the FE-OSV. No significant diffusion of Co atoms into the Alq3 layer is observed.

**Hysteretic behavior of the MR**

It has been known that MR depends sensitively on the measuring voltage ($V_{MR}$). Previous studies indicated that MR of LSMO/Alq$_3$/Co junctions (LSMO-OSV) increases with decreasing $V_{MR}$, reaching a maximum when $V_{MR}$ is close to zero.[5,16] For our FE-OSV junctions, its MR depends not only on $V_{MR}$, but also on the history of the voltage applied, giving rise to a strong hysteretic behavior. We characterize this hysteretic behavior by measuring MR($V_{MR}$) profile after applying a ramping voltage ($V_{MAX}$) that is larger than $V_{MR}$. (see Supplementary **Fig. S3** for the detailed measurement protocol).

As shown in **Fig. 2a**, for a FE-OSV containing an as-grown PZT layer, the MR (at $T = 11$K) is always negative and is the largest when $V_{MR}$ is near zero, which is similar to numerous previous studies on LSMO-OSV[2-6, 8-11]. The different features in the FE-OSV are: 1) the MR($V_{MR}$) profile is strongly affected by the initial voltage ($V_{MAX}$) applied; 2) a positive (negative) $V_{MAX}$ leads to a negative (positive) shift of the MR($V_{MR}$) profile along the $V_{MR}$ axis (**Fig. 2b** to **d**). This shift (ΔV) is closely tied to the hysteretic behavior of the PZT layer (so-called minor loops, see **Fig. 1e**) and increases with increasing $V_{MAX}$, as shown in **Fig. 2e**.

The hysteretic behavior of the MR was not observed in organic spin valves without the FE layer, including a conventional LSMO/Alq$_3$/Co organic spin valve (LSMO-OSV) and a LSMO/STO/Alq$_3$/Co organic spin valve (STO-OSV) in which the 5 nm PZT is replaced by 5 nm STO (see Suppl. Info. **Fig. S5** and **S7** for detailed MR loops and MR($V_{MR}$) profiles). This indicates that the hysteretic behavior of the MR in FE-OSV is tied to the presence of PZT. The distinctly different behavior between the FE-OSV and the PZT-free organic spin valves (LSMO-OSV and STO-OSV) also allow us to exclude the possibility of resistive bistability mechanism caused by the existence of the trap states or current conduct path inside the Alq$_3$ layer[20,21], since otherwise similar MR behavior should be observed in all three types of devices. We have also performed measurements on a LSMO/PZT/Co magnetic tunnel junction (FE-MTJ, see Fig.S6). While $V_{MAX}$ affects the MR($V_{MR}$) profile which confirms the results of Pantel *et al.*[28] and Valencia *et al.*,[30] it does not induce any shift of the MR($V_{MR}$) profile along the $V_{MR}$ axis. The comparison of the MR($V_{MR}$) profiles of the FE-OSV and the FE-MTJ suggests that the effective voltage generated by the dipole of the PZT (instead of the interfacial bonding) is responsible for the hysteretic behavior of the MR in the FE-OSV. The FE-MTJ does not show hysteretic behavior of MR because no direct dipole exists on the surface of the PZT due to the screening effect from the two metal electrodes in direct contact with the PZT layer.

A schematic model is illustrated in **Fig. 3** to explain this hysteretic behavior of MR in the FE-OSV. When a $V_{MAX}$ is applied to the junction, the polarization of the PZT gets modified. Due to the dipole moment created by the electric polarization of PZT, the effective voltage $V_{eff}$ applied on the Alq$_3$ layer equals $V_{MR}+\Delta$, where $\Delta$ is the vacuum level shift caused by the remnant dipole moment of PZT. When a positive (negative) $V_{MAX}$ is applied, $\Delta$ becomes larger (smaller), which explains why the MR ($V_{MR}$) profile can be shifted by $V_{MAX}$.

**Switching of the MR sign**

Remarkably, a reversal of the polarization of the PZT layer leads to a sign change of the MR in the FE-OSV. After measuring the MR and MR($V_{MR}$) profile (**Fig. 4 a-c**) for the FE-OSV device B (PZT layer in the as-grown state), we measured polarization-voltage dependence up to +/- 5 V. The measurement ended at -5.0 V in order to pole the PZT to the "down" state. The MR measurements of the FE-OSV device after this treatment are shown in **Fig. 4d**, **e**, **g**, and **h**. The shape of the MR($V_{MR}$) profile changes dramatically, as shown in **Fig. 4i**. In particular, the sign of the MR changes from negative (**Fig. 4a** and **b**) to positive (**Fig. 4d** and **g** and **h**) for a certain range of $V_{MR}$. A close correlation between the polarization of PZT and the sign of the MR can be identified by comparing the hysteretic behavior of the MR($V_{MR}$) profile and the possible minor polarization-voltage loop of the PZT, as illustrated in **Fig. 4f**, i.e. when the polarization of PZT is negative enough (more "down" state), the MR becomes positive.

The physical origin of MR sign of the LSMO/Alq3/Co junctions has been studied previously by Barraud *et al* using a scanning probe approach.[16] It was argued that firstly, the spin polarization alignment $P^*$ at the Co/Alq3 interface[16] is positive when electrons move from Co to Alq3, i.e. $P^*$(Alq3→Co)>0 where the arrow indicates the direction of the electric current;[3,22] secondly, the density of states for Alq3 at the Alq3/LSMO interface is spin polarized due to the coupling between the two materials, causing $P^*$(LSMO→Alq3)<0 when the Alq3 serves as a diffusive spacer.[16] Following these arguments, Barraud *et al* concluded that the sign of MR at small measurement voltages is determined by the sign of the product of $P^*$(Alq3→Co) and $P^*$(LSMO→Alq3); the result is negative. However, according to our observation, inserting a thin layer (5 nm) of as-grown PZT or STO between Alq3 and LSMO results in no sign change of MR, indicating no sign change of $P^*$(LSMO→Alq3). These results cannot be readily explained by the model of Barraud *et al*,[16] because there are no strong couplings between Alq3 and LSMO layers when they are separated by the PZT or STO layer.

In order to understand the negative MR in the LSMO-OSV or STO-OSV and the switching of MR sign in FE-OSV, we propose the following model based on the relative energy level alignment between Alq3 and LSMO. Here we consider mainly the hole transport because for Alq3 the energy difference between the HOMO and Fermi levels of the two metallic LSMO and Co electrodes are much smaller than that between the LUMO and the Fermi levels (See the Supp. Info., **Fig. S7**).[5,18,23] As shown in **Fig. 5a**, when the polarization of the PZT layer is pointing "up" (the as-grown state) or zero (the same case as for STO), the hole injection from the LSMO electrode for positive $V_{MR}$ is from the Fermi level (which lies in the spin majority band) of LSMO to the HOMO of Alq3, which is the same as in the LSMO-OSV. Therefore, $P^*(LSMO \rightarrow Alq_3)<0$, corresponding to a negative MR considering $P^*(Alq_3 \rightarrow Co)>0$ as discussed above.[3,22] When the polarization of the PZT layer is pointing "down", the HOMO of Alq3 is shifted up due to the dipole moment of PZT. Therefore, another state of the LSMO with opposite spin polarization may be accessible (**Fig. 5b**). Hence, $P^*(LSMO \rightarrow Alq_3)$ changes the sign and becomes positive, corresponding to a reversed, positive MR.

Our proposed model relies on two key assumptions: 1) the dipole moment of the PZT layer shifts the Alq3 HOMO level; and 2) the shift of the Alq3 HOMO level results in a shift of the initial state of LSMO for the hole injection between energy band of opposite spin polarizations. The first assumption was used by Schultz et al. to explain the MR sign reversal in a FeCo/Alq3/LiF/NiFe junction by proposing a shift of the HOMO of Alq3 due to the dipole moment of LiF layer.[18] The second assumption can be justified by the half metallicity of LSMO. As illustrated in **Fig. 5**, the conduction band of LSMO splits into spin majority and minority bands due to the exchange interaction, causing half metallicity because the Fermi level lies within the fully polarized spin majority band.[24–26] Therefore, when the HOMO of Alq3 is shifted up due to the reversal of dipole moment of PZT, the initial state of LSMO for hole injection may change to spin minority band with opposite spin polarization. <u>Note that this reversed dipole moment of PZT needs to be large enough to shift the HOMO level of Alq3 to '*reach*' the spin minority band of LSMO and obtain reversed positive MR values.</u>

Otherwise only negative MR values will be observed, as shown in **Fig. 4f** (also see Suppl. Info. **Fig. S8**).

Besides changing the energy level alignments, switching the electric polarization of the ferroelectric layer may also modify the coupling between the ferroelectric layer and the magnetic electrode, depending on the detailed electronic structure of the electrode and the nature of the electric polarization of the ferroelectric material.[27,28] These effects may also change the spin polarization at the interface between the ferroelectric material and the metal electrode.[27,28] However, both the magnetic structure of the LSMO and the spin polarization of the PZT are not expected to be affected very much by the electric polarization of the PZT, because of the robust magnetic properties of the $La_{0.7}Sr_{0.3}MnO_3$ with given composition (far from the metal-insulator phase boundary) and the large distance between the Mn site from LSMO and Ti sites from PZT at the interface.[28–30]

Another possible scenario involves the change of carrier type when the energy level alignment between LSMO and $Alq_3$ is changed: the carriers take the path of the HOMO (LUMO), i.e. hole (electron) transport in $Alq_3$ when the energy levels of $Alq_3$ is shifted "down" ("up") due to the "up" ("down") polarization of the PZT layer. This, however, contradicts our experimental observations because in the FE-OSV with the as-grown ("up" polarization) PZT, the MR is negative, the same as that in the STO-OSV or LSMO-OSV without any interfacial layers, suggesting that the carrier type in FE-OSV is most probably holes instead of electrons, as in the STO-OSV or LSMO-OSV.

**Conclusion and outlook**

The active control of the energy level alignment between the electrodes and the organic material, manifested here in the active control of the MR, not only carries promises for multistate control of organic spin-valve devices, but will also impact other organic electronic devices, in particular those applied in photovoltaics and solid state lighting. Specifically, the charge carrier injection efficiency of the organic light emitting diode (OLED) is determined by the relative alignment between the Fermi energy of the electrode and the energy levels of the organic material.[31] The charge collection efficiency in an organic photovoltaic (OPV) device also depends on the alignment of the energy levels of the acceptor organic material and the electrode.[32] Therefore, the realization of the active control of the level alignment using a ferroelectric interfacial layer demonstrated in this letter may also lead to successful optimization of other organic electronic devices by tailoring the energy landscape of the comprising materials using a tunable interfacial layer.

**Method Summary**

**Device fabrication:**

PZT, STO (5 nm) and LSMO (30 nm) thin films epitaxially grown on $SrTiO_3$ (001) substrate by pulsed laser deposition were fabricated into bottom electrodes using conventional wet-etch optical lithography. The $Alq_3$ (99.995%, Aldrich) layer (thickness: 65 nm) was deposited by thermal evaporation onto a room temperature substrate. The Co (10 nm) /Au (7 nm) was then deposited by thermal evaporation at substrate temperature of 280 K to complete the formation of the top electrode in a crossbar configuration. The device area is about 200 μm × 500 μm.

**Electrical characterization:**

Transport measurements were carried out using a Quantum Design Physical Property Measurement System (PPMS) combined with a Keithley 2400 source meter at $T$=11 K. Magnetic fields were applied in the plane of the thin film. The MR is defined as: $MR = (R_{antiparallel} - R_{parallel})/R_{parallel}$, where $R_{antiparallel}$ is the junction resistance in the antiparallel magnetic configuration and $R_{parallel}$ is the resistance at the parallel configuration.

**Acknowledgements**

This effort was supported by the National Basic Research Program of China (973 Program) under the grant No. 2011CB921800, 2013CB932901 and 2014CB921104, National Natural Science Foundation of China (91121002 and 11274071), Shanghai Municipal Natural Science Foundation (11ZR1402600), China Postdoctoral Science Foundation (2013M540321), the Wuhan National High Magnetic Field Center (WHMFCKF2011008) (MF, LY, YW, WY and JS). We also acknowledge the funding support of U.S. Department of Energy, Basic Energy Sciences, Materials Sciences and Engineering Division (DS, XSX, LJ, HNL, PCS, and TZW), and the U.S. Department of Energy, Basic Energy Sciences, Scientific User Facilities Division (XGZ and ZG), the US DOE grant DE-SC0002136 (HG and JS).

**Figure Legends:**

**Figure 1 | Structure of the organic spin valves and characterization of ferroelectricity in the epitaxial PZT. a**, Schematic structure of a Au/Co/Alq$_3$/PZT/LSMO organic spin valve (FE-OSV). **b**, AFM topography image of a PZT layer (5 nm in thickness) epitaxially grown on a LSMO (30 nm)/STO substrate. **c**, PFM (phase) response measured after successively switching the polarization of the PZT film by applying +2.5V and -2.5V on the tip with respective to the LSMO bottom electrode. Note that the protocol of voltage polarity is different in PFM measurements from that in resistance measurements. **d**, PFM image showing the polarization reversal by gradually increasing the applied voltage. The "up" ("down") arrow corresponds to the polarization pointing out of (into) the film surface. **e,** A typical PE loop for the FE-OSV (device A). The black and pink circles illustrate the "minor loops" corresponding to V$_{MAX}$ of +/- 0.5V and +/- 1.2V, respectively. The orange (blue) arrow indicates the direction of the ramp voltage: sweeping down (up) before the MR scans. **f**, TEM image for FE-OSV device. Different layers can be distinguished as labeled.

**Figure 2 | Hysteretic behavior of MR at *T*=11K. a**, Two MR scans in the as-grown state of a FE-OSV device (PZT in "up" polarization) with the same applied bias (V$_{MR}$) but different initial ramping voltage (V$_{MAX}$). Here LSMO is treated as anode. **b-d,** MR(V$_{MR}$) profiles taken at three different values of positive (orange) and negative (blue) V$_{MAX}$. ΔV represents the shift of MR (V$_{MR}$) profiles, and " **a'** " and " **a** " mark the positions at which the two MR loops in Fig. 2a are taken (V$_{MAX}$: +1.2V and -1.2V). **e**, Dependence of ΔV on V$_{MAX}$. ΔV for

both FE-OSV devices (A and B) increases with increasing $V_{MAX}$, while the three types of control devices (FE-MTJ, STO-OSV and LSMO-OSV) do not show any significant $\Delta V$.

**Figure 3 | Model of the hysteretic behavior in FE-OSV.** The relation between the effective bias on the spacer Alq$_3$ ($V_{eff}$), the applied bias at measurement ($V_{MR}$), and the vacuum level shift ($\Delta$) by the remanent dipole moments of the ferroelectric PZT layer, is $V_{eff} = V_{MR} + \Delta$.

**Figure 4 | Reversal of MR sign at *T*=11K. a** and **b,** MR scans for the "as-grown" state of the FE-OSV device B acquired at $V_{MR}$ =-0.3V / $V_{MAX}$= -0.5V and $V_{MR}$ =-0.3V / $V_{MAX}$= +0.5V, respectively. **c,** MR($V_{MR}$) profiles at $V_{MAX}$= +0.5V (orange) and -0.5V (blue). **d, e, g, and h,** MR scans taken when the polarization of the PZT is reversed. **f,** Comparison of the MR($V_{MR}$) profile and the polarization-voltage loop, showing the correlation between the sign of MR and the polarization of the PZT. The asymmetric regions of MR<0 and MR>0 in polarization-voltage loop indicate that the reversal of MR only occurs at higher degree of polarization in the "down" state of PZT. **i,** MR($V_{MR}$) profiles at $V_{MAX}$: +/- 0.5V when the polarization of the PZT is reversed. The red (blue) circles indicate the $V_{MR}$ values at which the MR loops (**a**, **b**, **d**, **e**, **g** and **h**) are taken with positive (negative) $V_{MAX}$.

**Figure 5 | Model of MR sign reversal in FE-OSV. a**, and **b** are the energy diagrams of the FE-OSV device when the electric polarization of the PZT is "up" and "down" respectively. The white circles show the injected holes in the device. The blue (red) arrows show the spin

polarized holes injected from the majority (minority) band of LSMO.

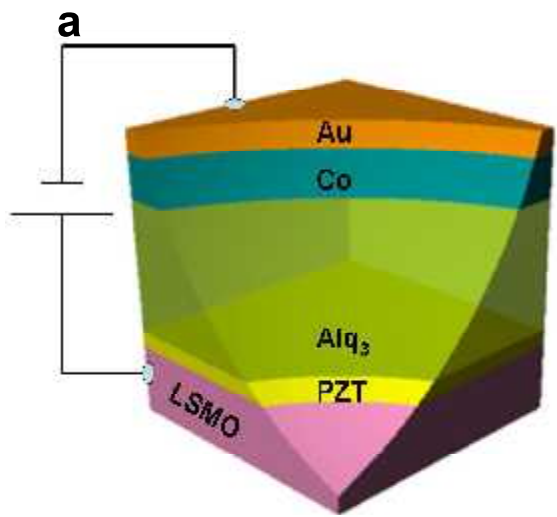
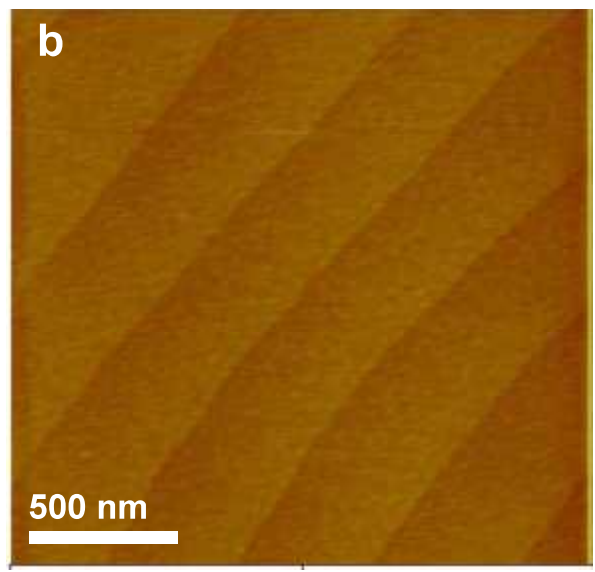
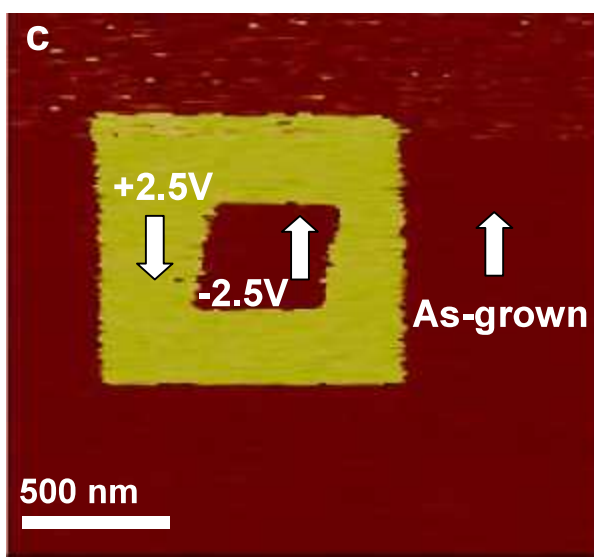
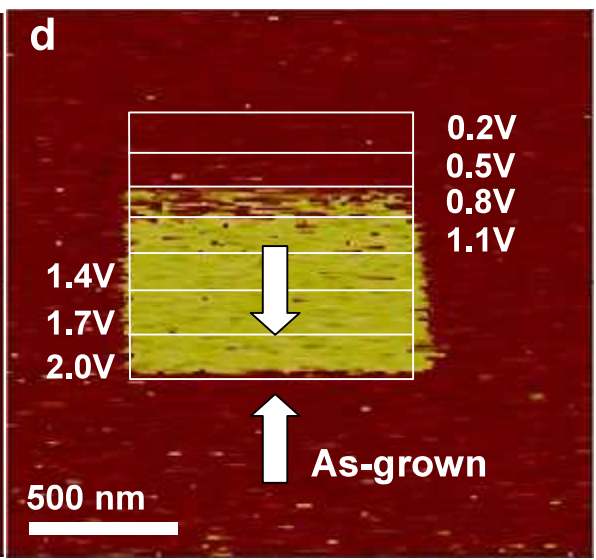
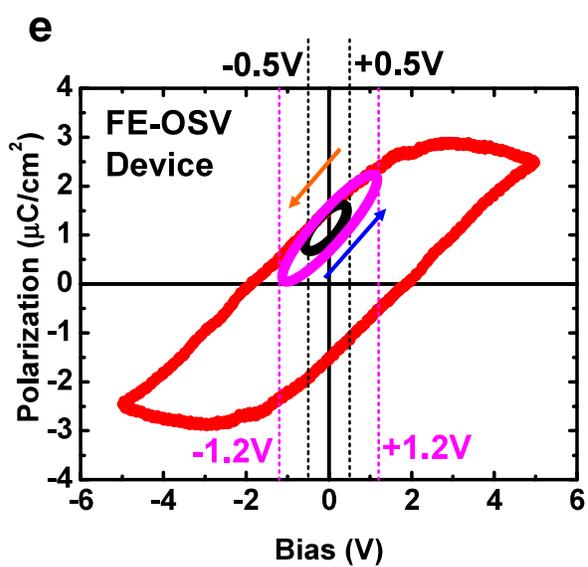
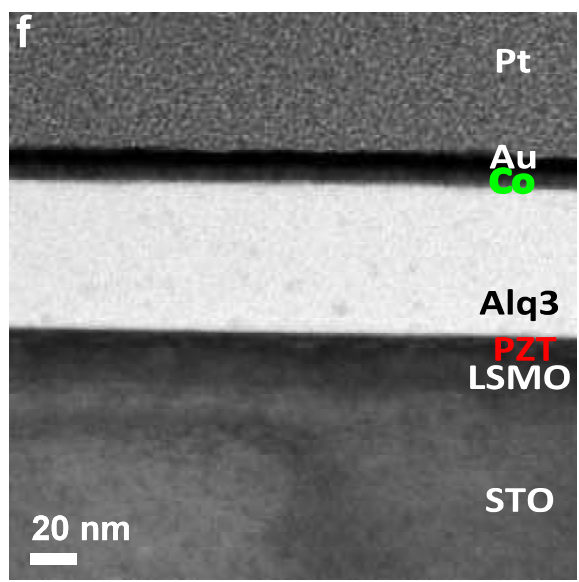

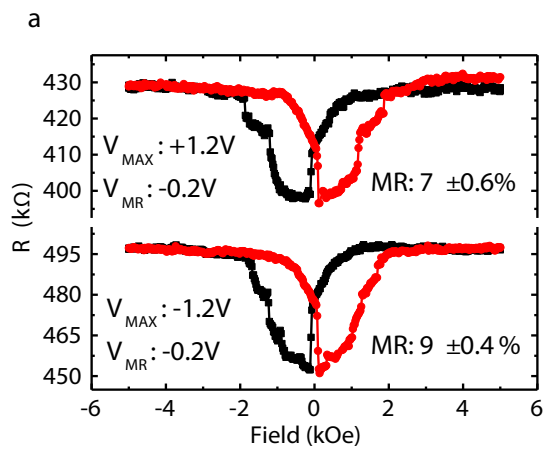
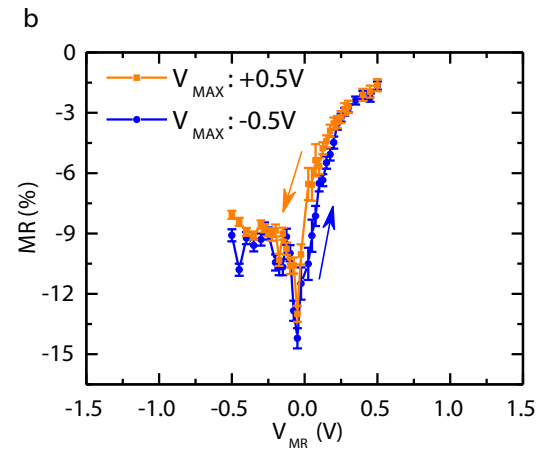
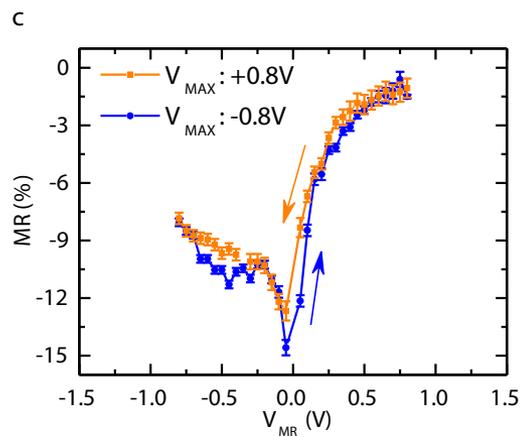
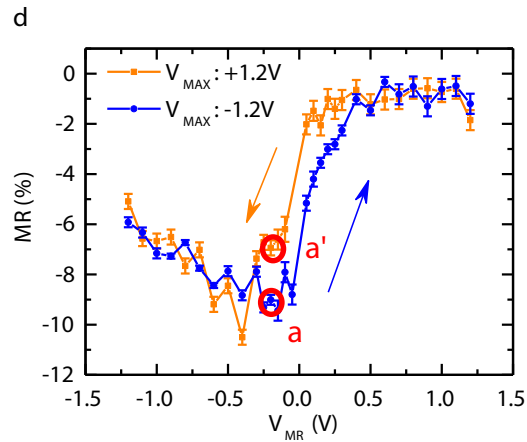

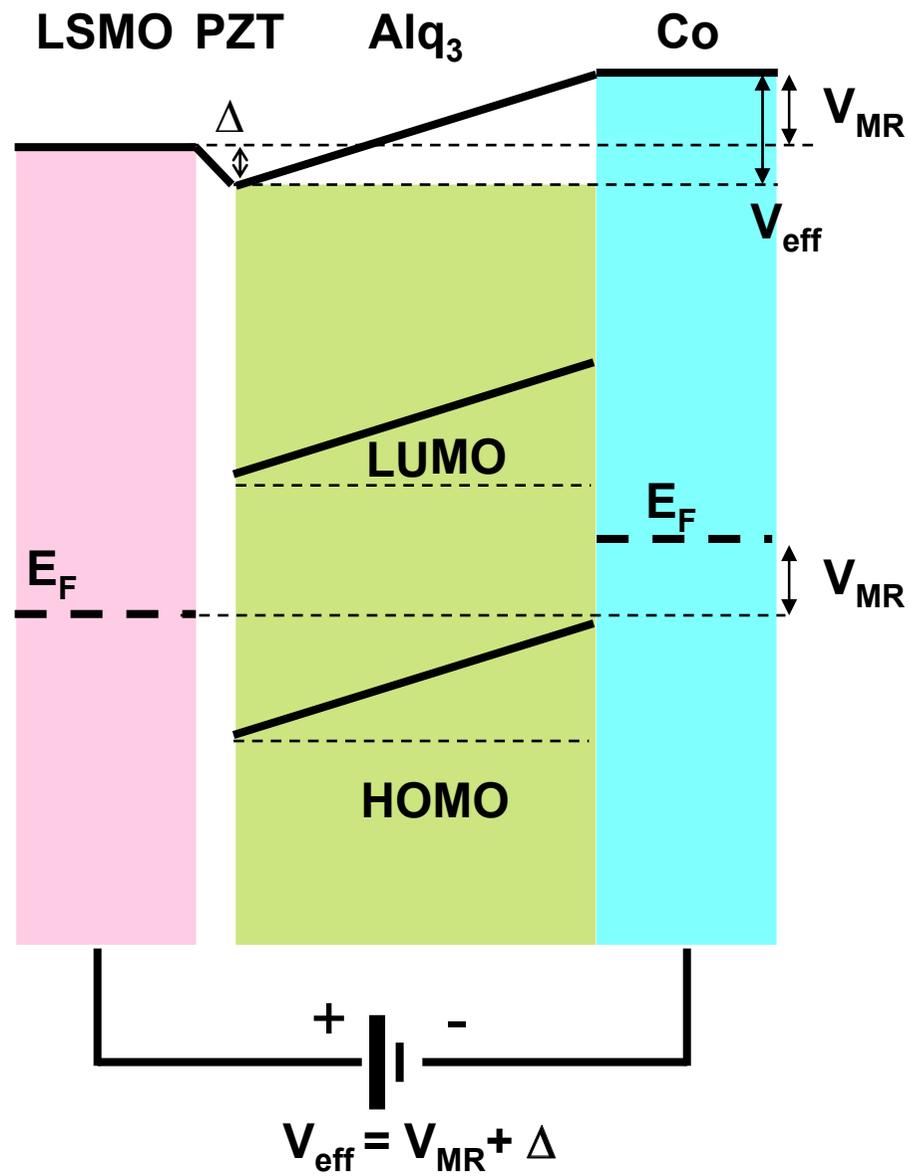

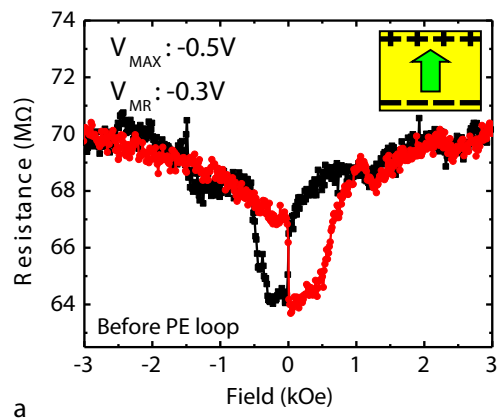a

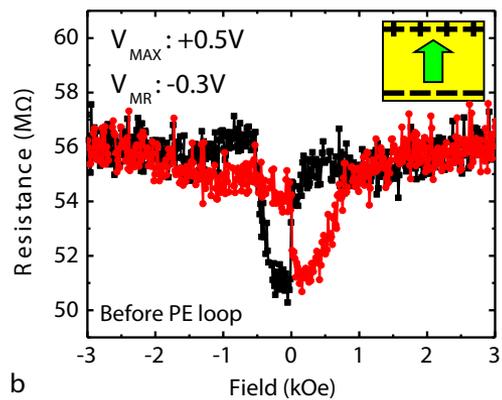b

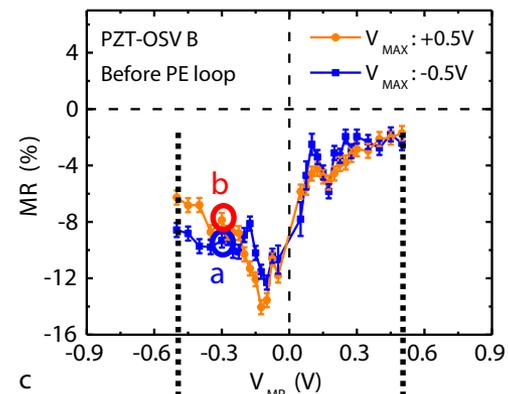c

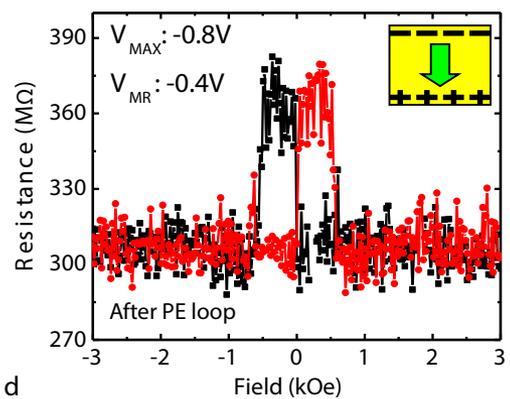d

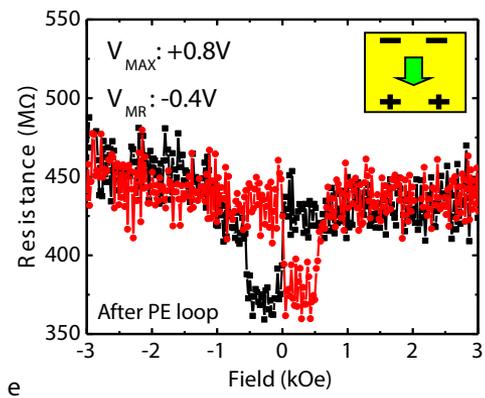e

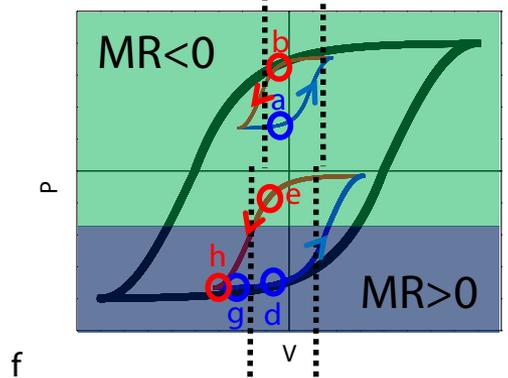f

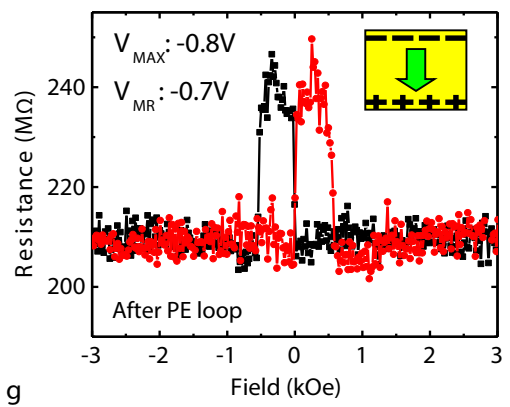g

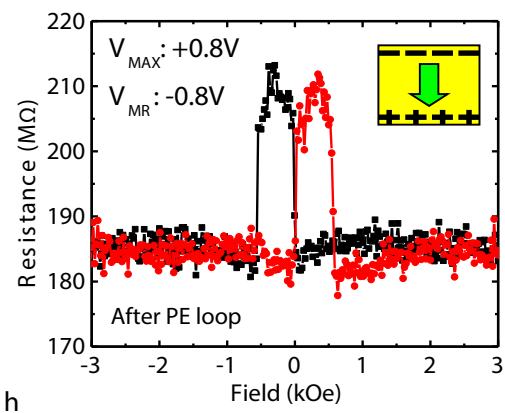h

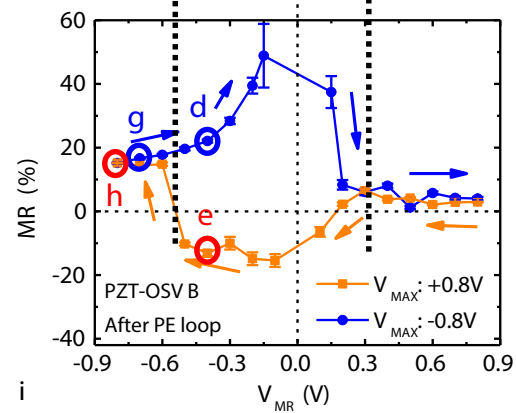i

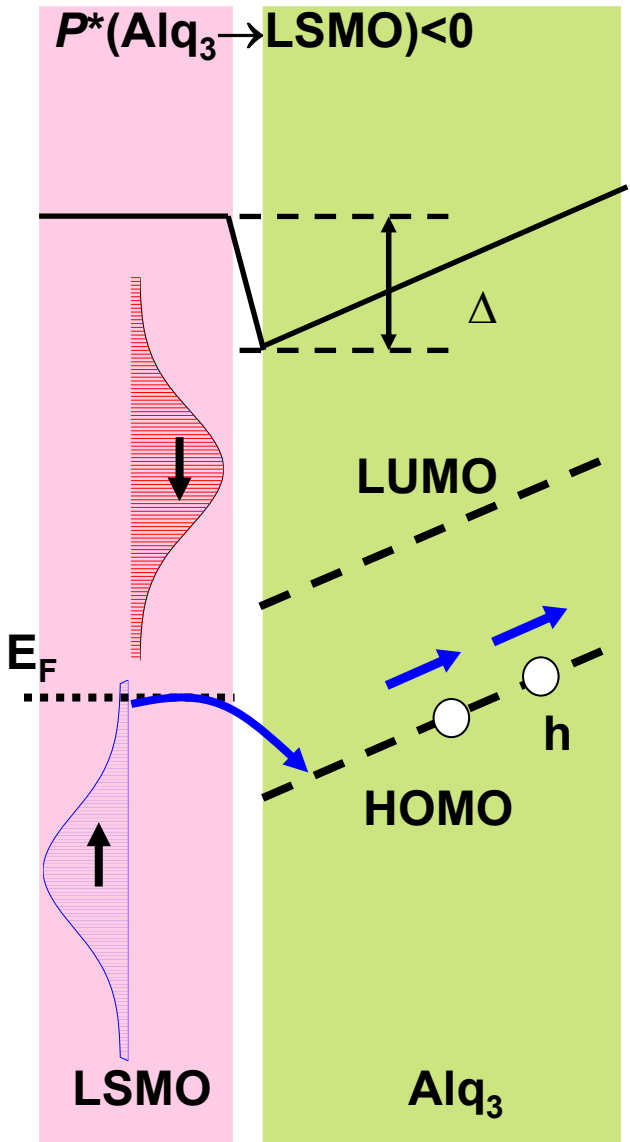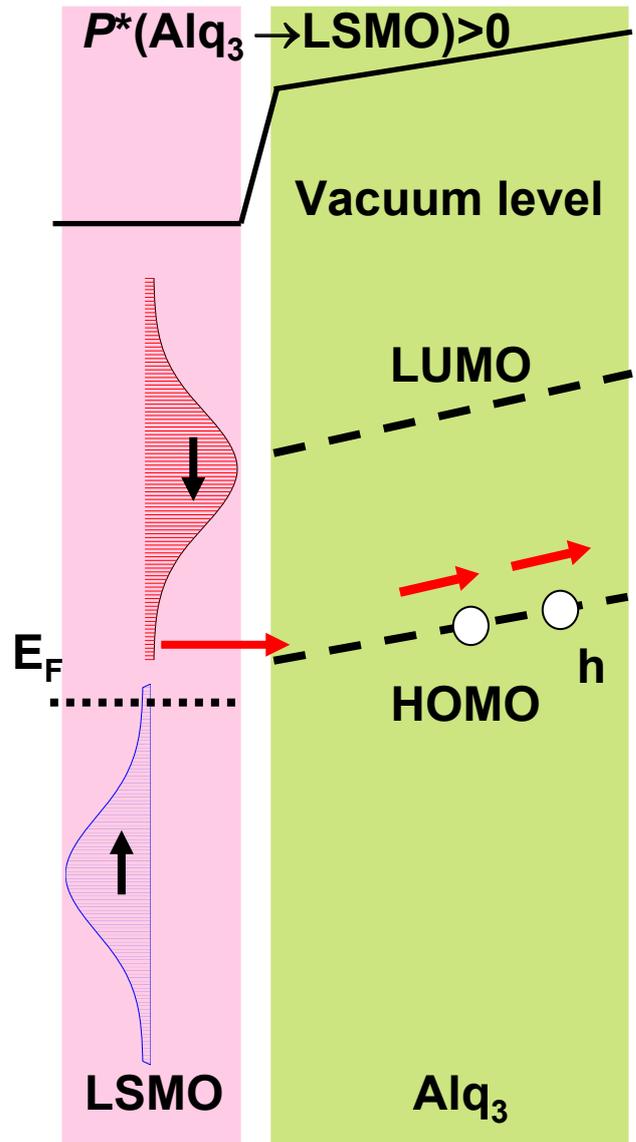